\documentstyle[12pt]{article}

\newcommand{\BEQ}{\begin{equation}}
\newcommand{\BEA}{\begin{eqnarray}}
\newcommand{\EEQ}{\end{equation}}
\newcommand{\EEA}{\end{eqnarray}}
\newcommand{\eps}{\epsilon}       
\newcommand{\vph}{\varphi}        
\newcommand{\rar}{\rightarrow}    

\newcommand{\zeile}[1]{\vskip #1 \baselineskip}  

\newcommand{\Gll}{C(r_{\parallel}, r_{\perp};T_c)}
\newcommand{\GlT}{C(\vec l;T)}
\newcommand{\GlTc}{C(\vec l;T_c)}
\newcommand{\GlTco}{C_1(\vec l;T_c)}
\newcommand{\FB}[3]{\hbox{#1}_{#2}(#3)} 

\newcommand{\appsection}[1]{\setcounter{equation}{0} \section*{Appendix #1}}
\renewcommand{\theequation}{\thesection.\arabic{equation}}

\newcommand{\ee}{{\hbox{\rm e}}}

\catcode`\@=11
\def\numberbysection{\@addtoreset{equation}{section}
        \def\theequation{\thesection.\arabic{equation}}}
\numberbysection

\begin{document}

\begin{titlepage}
\null
\vskip 4cm
\begin{center}
\vskip 1.in
{\Large \bf Exact Correlation Function at the Lifshitz Points}
\zeile{1.2}
{\Large \bf  of the Spherical Model}
\footnote{Supported in part by the Swiss National Science Foundation}
\vskip 0.5in
 Laurent Frachebourg and Malte Henkel
 \\[.3in]
{\em D\'epartement de Physique Th\'{e}orique,
     Universit\'e de Gen\`eve \\
     24  quai Ernest Ansermet,
     CH - 1211 Gen\`eve 4, Switzerland}
\zeile{2}
{\bf UGVA-DPT 1992/12-796}
\end{center}
%
%
\zeile{3}

The spin-spin correlation function of the spherical model being
precisely at an anisotropic Lifshitz point of arbitrary order is calculated
exactly. The results are in agreement with scaling. The scaling
function is shown to be universal.
The direction-dependent long-range correlations may
change from ferromagnetic to antiferromagnetic
behaviour and back as the dimension
is varied. The form of the scaling function is compared to predictions
following from local scale invariance for strongly
anisotropic critical systems.

\end{titlepage}

\newpage

%
%
\section{Introduction}

In magnetism, a Lifshitz point (of first order)
is defined as the meeting point of the
transitions between a paramagnetic, a ferromagnetic and an ordered
incommensurate phase \cite{Horn75}.
Lifshitz points have been observed experimentally,
a familiar example being MnP \cite{Becc80}.
They can be realized in lattice models by considering anisotropic
competing interactions extending beyond the conventional nearest-neighbour
interactions. Models of this kind have been investigated extensively, see
\cite{revselke} for a recent review.
A well-studied example is provided by the ANNNI model \cite{Selk80}. While the
ANNNI model only contains
next-to-nearest neighbour interactions, the extension to more
general interaction types with Hamiltonian
${\cal H} = \sum_{\vec{a}} {\cal H}(\vec{a})$ where
\BEQ \label{HHH}
{\cal H}(\vec{a}) = -J \left( \sum_{j=1}^{d} s_{\vec{a}}
s_{\vec{a}+\vec{e}_{j}}
+ \sum_{j=1}^{m} \sum_{i=1}^{n} \kappa_n s_{\vec{a}}
s_{\vec{a}+(i+1)\vec{e}_{j}} \right)
\EEQ
has
been studied at length as well. Here $s_{\vec{a}}$ are the spin variables
at site $\vec{a}$ of a hypercubic lattice of dimension $d$,
$m$ is the number of directions in
which the competing interactions are taken,
$n$ is the number of interacting neighbours in
these directions and $\kappa_i$ parametrize the competing
interaction terms.  We take $J$ to be positive.

The definition of Lifshitz points \cite{Selk77,Nico76}
is quite analogous to the definition
of multicritical points. The phase diagramm of the model eq.~(\ref{HHH})
can be quite
complex and contains both ordered ferromagnetic and helicoidal phases.
A Lifshitz point of first arises when these two ordered phases meet with
the disordered one \cite{Horn75}. If several of
the $\kappa_i$ are non-vanishing,
the phase diagram may contain a line of Lifshitz points of first order.
This line terminates in a Lifshitz point of second order. Lifshitz points
of higher order can be defined analogously.

The anisotropies introduced in the lattice
model eq.~(\ref{HHH}) may survive in the thermodynamic limit.
The critical behaviour of the correlation functions in
a system being exactly at
the Lifshitz point becomes dependent on
the space direction. The correlation function
$C(\vec r_{\parallel},\vec  r_{\perp})$ depends on whether
correlations are studied along those $d-m$ directions
with only nearest neighbour interactions
(referred as to $\perp$) or those $m$ directions where
competing interactions are present (referred to as
$\|$). At criticality, one has \cite{Horn75}:
\BEA
C(0,\vec  r_{\perp})     & \sim &
 r_{\perp}^{-\left(d-d_-+\eta_{\perp}\right)}  \nonumber\\
C(\vec r_{\parallel},0) & \sim &
r_{\parallel}^{-({d-d_-\over \theta}+\eta_{\parallel})} \label{eq:AxScal}
\EEA
where $d_-$ is the lower critical dimension and $\theta$ is defined below.
This defines the
direction-dependent exponents $\eta_{\|,\perp}$ which for Lifshitz points
of first order are also referred to as
$\eta_{\ell 4,\ell 2}$, respectively, in the literature.
Similarly, two types of correlation length
$\xi_{\parallel},\xi_{\perp}$ are defined and are
direction-dependent as well:
\BEQ
\xi_{\parallel}\sim (T-T_c)^{-\nu_{\parallel}},
\qquad \xi_{\perp}\sim (T-T_c)^{-\nu_{\perp}}
\EEQ
while the critical exponents $\alpha,\, \beta,\,
\gamma$ can be defined as usual from the
specific heat, the order parameter and the susceptibility.
The scaling relation among the exponents for isotropic systems
are replaced by anisotropic scaling relations \cite{Horn75}
\BEA
2-\alpha & = & m\nu_{\parallel}+(d-m)\nu_{\perp}\nonumber\\
\gamma & = & (2-\eta_{\perp})\nu_{\perp}=
\left({2\over \theta}-\eta_{\parallel}\right)\nu_{\parallel}\label{scal}
\EEA
where
\BEQ
\theta={\nu_{\parallel}\over \nu_{\perp}}
\EEQ
is the anisotropy exponent.
Equations~(\ref{scal}) replace the conventional
scaling relations involving $\nu$ and $\eta$.
Consequently, there are three independent critical
exponents which describe the leading bulk critical behaviour.

The strong anisotropy of a system being at the Lifshitz
point leads, via standard renormalization group arguments \cite{Horn75},
to the following well-known scaling of the
correlation function
\BEQ
C(\lambda\vec r_{\parallel},\lambda^{1/\theta}
\vec  r_{\perp})=\lambda^{-({d-d_-\over \theta}+\eta_{\parallel})}
C(\vec r_{\parallel},\vec  r_{\perp})\label{scal2}\, .
\EEQ
This is equivalent to the scaling form:
\BEQ
C(\vec r_{\parallel},\vec  r_{\perp})\sim
   r_{\perp}^{-(d-d_-+\eta_{\perp})}
\Phi\left({ r_{\parallel}\over  r_{\perp}^{\theta}}\right)\label{ff}
\EEQ
which defines the scaling function $\Phi(x)$ where
\BEQ
x={ r_{\parallel}\over  r_{\perp}^{\theta}}
\EEQ
is the scaling variable. Note that any attempt to calculate $\Phi$ from a
lattice model requires that the scaling limit $r_{\|}\rar\infty$,
$r_{\perp}\rar\infty$ such that $x$ is kept fixed has to be taken.

In this paper, we consider Lifshitz points of arbitrary
order in the spherical model with additional competing interactions in
$m$ directions. In the literature, this
system is known as the ANNNS model \cite{revselke,Selk77}
or the R-S model \cite{stanley}. Throughout this paper, we restrict attention
to anisotropic Lifshitz points, that is we only consider the situation
where $1 \leq m \leq d-1$. Although
this model is of no direct experimental relevance,
it may provide useful insight since all physical
quantities of interest can be evaluated exactly. In this respect, the
spherical model has been quite a useful tool in providing explicit checks
on general concepts in critical phenomena,
see \cite{Barb73,Joyc76,Baxt82,Sing84,Bran92,Coni89,Khor92}.
The critical exponents of the ANNNS model
are for dimensions between
the lower critical dimension $d_{-}$ and the upper critical dimension
\BEQ
d_{+}= d_{-} +2 = 4 + m - m/L
\EEQ
given by \cite{Selk77}
\BEQ
\eta_{\parallel}=\eta_{\perp}=0 \qquad ,
\qquad \nu_{\parallel}={\gamma\over 2L} \qquad ,
\quad \nu_{\perp}={\gamma\over 2}\nonumber
\EEQ
\BEA
\gamma & = & {2L\over (d-2-m)L+m}\nonumber  \\
\alpha & = & {m+L(d-4-m)\over m+L(d-2-m)}
\EEA
for a Lifshitz point of order $L-1$.
Here we obtain the exact scaling function $\Phi(x)$
in eq.~(\ref{ff}) for anisotropic Lifshitz points of arbitrary order.

Finding the exact scaling function of a system with a strongly anisotropic
scaling, see eq~(\ref{scal2}), is of interest by itself. In particular,
we shall find regions of values of the scaling variable $x$, where the
long-range correlations become antiferromagnetic.
However, our main motivation for undertaking this work is
as follows. Given the fact that local scale invariance
has lead to an enormous increase of understanding the
critical behaviour of static, isotropic systems using
conformal invariance, at least in two dimensions
(see e.g. \cite{konform}), one may
wonder whether at least some of these ideas might
be useful in more general situations.
In fact, Cardy \cite{Cardy} had proposed to use conformal
invariance in the context of critical
dynamics, starting from the hypothesis of dynamical scaling
of the time-dependent correlation function
$\langle \phi(r,t)\phi(0,t)\rangle = t^{-2x/z}\Phi(r^{2}/t)$.
He considered time-dependent systems in two space dimensions
where the static system by itself is conformal
invariant, this is, at the critical point. Using conformal
transformations, the problem is mapped
from the two-dimensional infinite plane to a strip
of finite width and it is argued, since the strip is finite
that it were permissible to use van Hove theory.
For a purely relaxational dynamics without any macroscopic
conservation law, this leads to \cite{Cardy}
\BEQ
\Phi(y)=\hbox{e}^{-\mu y} \label{ff1}
\EEQ
where $\mu$ is a non-universal constant. Note that this result
is apparently independent of the dynamical exponent $z$.

On the other hand, for $z=2$ but for an arbitrary number of space dimensions,
the global scale invariance eq.~(\ref{scal2}) can be generalized to
a local one \cite{malte1}. Then the coordinate
transformations to be considered are those given by the
Schr\"odinger group. Then for example the two-point time-delayed correlation
function $\langle \phi(r,t)\phi(0,0)\rangle = t^{-2x/z} \Phi(r^{2}/t)$
is fixed completely where $\Phi(y)$ is in turn
given by eq.~(\ref{ff1}) \cite{malte1},
but now without the restriction to two space dimensions and
without having to appeal to van Hove theory. We stress that $z=2$ does not
need imply that the system is described by van Hove (mean field)
theory, the best-known example probably being the one-dimensional
Ising model with Glauber dynamics \cite{Glau63}.
The form of the three-point correlation functions was also found.

We consider the spherical model at a Lifshitz point
of order $L-1$ as a convenient tool to test these general
ideas using the following analogy.
In critical dynamics, we have $d$ space dimensions
and one time direction, the rescaling of which is described by
the dynamical exponent $z$. This situation can be
mimicked by considering either the case of just $m=1$ direction
with additioned competing interactions which
leads to the analogy $z\equiv \theta={1\over L}$,
or the case $m=d-1$ which would be analogous to $z\equiv {1\over \theta}=L$.
As we shall show, there exist examples confirming the hypothesis of local
scale invariance.

The paper is organized as follows. In section 2 we give the general
procedure to calculate the correlation function. The case of a Lifshitz point
of first order is described in section 3, while Lifshitz points of higher
order are treated in section 4. In section 5 we compare the exact results
obtained with the expectations from local scaling and conclude.

\section{Correlation functions}

The correlation function for the (mean) spherical model is defined as
\BEQ
C_N\left(\vec l,T\right)\equiv
<\sigma_{\vec 0}\sigma_{\vec l}>_N-<\sigma_{\vec 0}>_N^2
\EEQ
As is well known, see e.g. \cite{Barb73,Joyc76,Baxt82,Khor92}, it is
given in the thermodynamic limit $(N\rightarrow \infty$) by
\BEQ \label{CorrFunc}
\GlT={kT\over(2\pi)^d}\int_{-\pi}^{\pi}\ldots
\int_{-\pi}^{\pi}d\phi_1...d\phi_d \,
\cos\left(\vec{l}\,\vec{\phi}\right)\over \tilde J(0)\zeta
-\tilde J(\vec \phi)
\EEQ
where $\tilde J(\vec \phi)$ is the Fourier transform of the exchange integral
\BEQ
\tilde J(\vec \phi)=\sum_{\vec l}J(\vec l)
\exp \left({i\vec{l}\, \vec{\phi}}\right)
\EEQ
and $\zeta$ is given by the spherical constraint. It can be shown that
$\zeta=1$ for $T\leq T_c$. We are interested in the
behaviour of eq.~(\ref{CorrFunc}) at the critical
point where $\zeta=1$ and we restrict
ourselves to this case throughout the paper.
In fact, the case we are going to consider
is technically the hardest one and
all other situations of physical interest
are easily obtainable from our results.

For a $m$-axial Lifshitz point of order $L-1$,
$\tilde J(\vec \phi)$ is, see \cite{Selk77,Nico76}
\BEQ
\tilde{J}(\vec \phi)=2J\left(\sum_{j=1}^d\cos\phi_j+\sum_{j=1}^m
\sum_{i=1}^n\kappa_i \cos((i+1)\phi_j)\right)
\EEQ
which can be expanded
\BEQ
\tilde{J}(\vec \phi)=2Jd+m\left(\sum_{i=1}^{n}\kappa_i\right)
-{1\over 2}\sum_{j=m+1}^d\phi_j^2-c_L\sum_{j=1}^m\phi_j^{2L}+...
\EEQ
Consider for a moment the case $n=2$. A line of first-order Lifshitz points is
obtained for
$\kappa_2=-{(1+4\kappa_1)/9}$ where $\kappa_1>-2/5$ is a free parameter and
$c_2={(2+5\kappa_1)/6}>0$.
A second-order Lifshitz point is found if $\kappa_1=-{2/5}$ and
$\kappa_2={1/15}$ with $c_3={1/30}$.
The phase diagram is
known exactly \cite{Horn76} and we do not repeat this calculation here.
Similar results could be
obtained without much effort for $n$ arbitrary and yield explicit
expressions for $c_L$.
If the $\kappa_i$'s were chosen
such that $c_L$ would become negative,
the correlation function becomes modulated
and the simple expansion around the assumed ground state given by
$\vec{\phi}=0$ is no longer applicable.

The evaluation of the correlation function closely follows techniques which
go back at least to the classic work
\cite{Montroll} on random walks on the lattice.
As shown in \cite{Montroll}, for $|\vec l|$ big enough
the principal contribution to the integrals
comes from $|\vec{\phi}| \sim  0$, and the leading
singular behaviour of $\GlTc$ is contained in
\BEQ
\GlTco={kT_c\over 2J(2\pi)^d}\int_{-\pi}^{\pi}\ldots
\int_{-\pi}^{\pi}d\phi_1\ldots d\phi_d
{\cos(\vec{l}\,\vec{\phi})\over \sum_{j=m+1}^d
{1\over 2}\phi_j^2+c_L\sum_{j=1}^m\phi_j^{2L}}.
\EEQ
It is here that the scaling limit mentioned in the introduction
is taken. In this limit, it is enough to restrict attention to $C_1$,
since all other terms contributing to $C$ can be made arbitrarily
small \cite{Montroll}. In the sequel, we always suppress the
distinction between $C$ and $C_1$. Using the identity
\BEQ
a^{-1}=\int_0^{\infty}e^{-au}du
\EEQ
we can rewrite $\GlTc$ as
\BEQ
\GlTc={kT_c\over 2J(2\pi)^d} \int_0^{\infty}du F(u)
\EEQ
with
\BEQ
F(u)=\Re\left\{ \prod_{j=1}^m
\left[\int_{-\pi}^{\pi}\exp\left(i\phi_j l_j-c_L\phi_j^{2L}u\right)
d\phi_j\right]
\prod_{j=m+1}^d \left[\int_{-\pi}^{\pi}\exp
\left(i\phi_j l_j-{1\over 2}\phi_j^2u\right)d\phi_j \right]\right\}
\EEQ
where $\Re$ denotes the real part.
An Abelian theorem for the Laplace transform \cite{Montroll}
states that if $F(u)$ is analytic,
the behaviour of $\GlTc$ is determined
by the behaviour of $F(u)$ at $u\rightarrow \infty$.
For $u\gg1$ we can therefore replace the range of
integration $(-\pi,\pi)$ on the $\phi_j$ integrals by $(-\infty,\infty)$.

The correlation function now reads
\BEA
\GlTc & = & {kT_c\over 2J(2\pi)^d}\int_0^{\infty}du\left(
\Re \left\{\prod_{j=1}^m \left[\int_{-\infty}^{\infty}
\exp\left(i\phi_j l_j-c_L\phi_j^{2L}u\right)d\phi_j \right]\right\}\right.
\nonumber\\
       &   & \left.\mbox{}\times \left({2\pi\over u}\right)^{{d-m\over 2}}
\exp\left(-{1\over 2u}\sum_{j=m+1}^dl_j^2\right)  \right)
\EEA
Expanding the cosine and integrating term by term we get
\BEQ
\int_{-\infty}^{\infty} \ee^{-c_L\phi_j^{2L}u}
\cos(l_j\phi_j)d\phi_j={1\over L}\left({1\over c_Lu}\right)^{1\over 2L}
\sum_{k=0}^{\infty}{(-1)^k\over (2k)!}
\Gamma \left({k\over L}+{1\over 2L}\right)
\left({l_j^2\over (c_Lu)^{1\over L}}\right)^k
\EEQ
In the sequel we specialize to the choice
\BEA
 r_{\parallel} &=& l_1 \neq 0\;\; ,\;\; l_i=0 \; (i=2,..,m), \nonumber \\
 r_{\perp}^2 &=& \sum_{j=m+1}^{d} l_j^2
\EEA
and the correlation function becomes
\BEA
\lefteqn{ \Gll = {kT_c\over 2J(2\pi)^{{d+m\over 2}}}
\left({1\over L}\right)^m\Gamma\left({1\over 2L}\right)^{m-1}
\left({1\over c_L}\right)^{m\over 2L} }  \\
&& \times \int_0^{\infty}du \left\{ u^{{m\over 2}\left(1-{1\over L}\right)
-{d\over 2}}\exp\left(-{ r_{\perp}^2\over 2u}\right)
\left(\sum_{k=0}^{\infty}{(-1)^k\over (2k)!}
\Gamma \left({k\over L}+{1\over 2L}\right)
\left({ r_{\parallel}^2\over (c_Lu)^{1\over L}}\right)^k\right)
\right\} \nonumber
\EEA
The integral converges if
\BEQ
d>2+m-{m\over L}=d_{-}
\EEQ
and the sum is absolutely convergent. Exchanging
the sum with the integral, we thus obtain
an exact expression for the correlation function
\BEA
\Gll & = & {kT_c\over 2J(2\pi)^{{d+m\over 2}}}
\left({1\over L}\right)^m\Gamma\left({1\over 2L}\right)^{m-1}
\left({1\over c_L}\right)^{m\over 2L}
\left({ r_{\perp}^2\over 2}\right)^{1+{m\over 2}
\left(1-{1\over L}\right)-{d\over 2}}\nonumber \\
     &  & \mbox{}\times \sum_{k=0}^{\infty}{(-1)^k\over (2k)!}
\Gamma \left({k\over L}+{1\over 2L}\right)
\Gamma \left({k\over L}+{d-d_-\over 2}\right)
\left( \left( \frac{2}{c_{L}} \right)^{1/L}
 \frac{ r_{\parallel}^2}{ r_{\perp}^{2/L}} \right)^k \, \nonumber \\
& &
\EEA
We remark that the series is absolutely convergent
on the whole real axis and that the correlation function is of the form
\BEQ \label{GenRes}
\Gll \sim A(L,m,d)  r_{\perp}^{-(d-d_-)}
\Phi\left(\left(\frac{2}{c_L}\right)^{\theta/2}
{ r_{\parallel}\over  r_{\perp}^{\theta}};L,m,d\right)
\EEQ
where $A(L,m,d)$ is a constant and
$\Phi(x;L,m,d)$ the desired scaling function.
This is in agreement with the expected scaling
form eq.~(\ref{scal2}) and serves as a useful check of our calculation.
Furthermore, since the scaling form of the
correlation function only depends
through the constant $c_L$ on the details of the
lattice structure, we verify the universality of the scaling function.

While this already an exact and complete answer to our problem,
it is useful to rewrite this result in a more handy
form. This will be done in the next sections.
The general strategy is as follows.
We use first the identity
\BEQ
(2k)!=k!(2\pi)^{-{L\over 2}}2^{2k+{1\over 2}}L^k
\prod_{n=0}^{L-1}\Gamma\left({k\over L}+{1\over 2L}+{n\over L}\right)
\label{factor}
\EEQ
and the correlation function finally becomes
\BEQ \label{eq:Skalenform}
\Gll = B(L,m,d)  r_{\perp}^{-(d-d_{-})}
\Xi\left(L,\frac{d-d_{-}}{2};{2^{{1\over L}}\over 4Lc_L^{{1\over L}}}
{ r_{\parallel}^2\over  r_{\perp}^{{2\over L}}}\right)
\EEQ
where the constant $B(L,m,d)$ is given by
\BEQ
B(L,m,d)={kT_c\over J(2\pi)^{{d+m-L\over 2}}}
\left({1\over L}\right)^m\Gamma\left({1\over 2L}\right)^{m-1}
\left({1\over c_L}\right)^{m\over 2L}
2^{{d-d_-\over 2}-{3\over 2}}
\EEQ
All properties of $\Gll$ are contained within the series
\BEQ
\Xi(L,a;x)=\sum_{k=0}^{\infty}{(-1)^k\over k!}
{\Gamma \left({k\over L}+a\right)\over
\prod_{n=1}^{L-1}\Gamma\left({k\over L}+{1\over 2L}+{n\over L}\right)}x^k
\label{eq:XiReihe}
\EEQ
which we proceed to study in the following.
It is sufficient to consider only those cases where $0 < a \leq 1$, since
the other cases can be found from the recursion
\BEQ \label{XiRekurs}
\Xi(L,a+1;x) = a \Xi(L,a;x) + \frac{1}{L} x \frac{\partial}{\partial x}
\Xi(L,a;x)
\EEQ
In the sequel, we shall use the abbreviation
\BEQ \label{eq:adef}
a = \frac{d}{2} - \frac{m}{4} -1 = \frac{d - d_{-}}{2}
\EEQ
For $a=0$, the model is at the lower critical dimension $d_{-}$, while for
$a=1$, it is at the upper critical dimension $d_{+}$. For the convenience of
the
reader, we give in table~1 the values of $a$ if both $d$
and $m$ are integers for the
cases $L=2$ and $L=3$.
\begin{table}[htb]
\begin{center}
\begin{tabular}{|c|cccccc|} \hline
$L=2$ & \multicolumn{6}{c|}{$m$} \\ \hline
$d$ & 1 & 2 & 3 & 4 & 5 & 6  \\ \hline
3 & ${1}/{4}$ & $0$ & & & & \\
4 & ${3}/{4}$ & ${1}/{2}$ & ${1}/{4}$ & & & \\
5 & ${5}/{4}$ & $1$ & ${3}/{4}$ & ${1}/{2}$ & & \\
6 & ${7}/{4}$ & ${3}/{2}$ & ${5}/{4}$ & $1$ & ${3}/{4}$ & \\
7 & ${9}/{4}$ & $2$ & ${7}/{4}$ & ${3}/{2}$ & ${5}/{4}$ & $1$ \\ \hline
\end{tabular} \end{center} ~ \\ ~
\begin{center}
\begin{tabular}{|c|cccccccc|} \hline
$L=3$ & \multicolumn{8}{c|}{$m$} \\ \hline
$d$ & 1 & 2 & 3 & 4 & 5 & 6 & 7 & 8\\ \hline
3 & $1/6$ & & & & & & &  \\
4 & $2/3$ & $1/3$ & 0 & & & & & \\
5 & $7/6$ & $5/6$ & $1/2$ & $1/6$ & & & & \\
6 & $5/3$ & $4/3$ & $1$ & $2/3$ & $1/3$ & & & \\
7 & $13/6$ & $11/6$ & $3/2$ & $7/6$ & $5/6$ & $1/2$ & & \\
8 & $8/3$ & $7/3$ & $2$ & $5/3$ & $4/3$ & $1$ & $2/3$ & \\
9 & $19/6$ & $17/6$ & $5/2$ & $13/6$ & $11/6$ & $3/2$ & $7/6$ & $5/6$ \\ \hline
\end{tabular} \end{center}
\caption{Some values of the parameter $a=\frac{1}{2}\left( d - d_{-}\right)$
for $L=2$ and $L=3$ as a function of $d$ and $m \leq d-1$.}
\end{table}

If $d$ and $m$ are integers, $\Xi(L,a;x)$
can be reexpressed in terms of well-known transcendental functions. We shall
derive these expressions for $L=2$ and $L=3$ below.
In particular, we shall be interested in deriving the
behaviour of the correlation function for large values of the
scaling variable $x$ and we shall obtain explicit
expressions for any $L$.

\section{Lifshitz points of first order}

We first study the case of a conventional Lifshitz point, also referred
to as a Lifshitz point of first order
\cite{Selk77}. This corresponds in the above equations to have $L=2$.
As we have seen above, the correlation function becomes
\BEQ
\Gll = B(2,m,d)  r_{\perp}^{-(d-d_{-})}
\Psi\left( \frac{d-d_{-}}{2} , \sqrt{ \frac{1}{32 c_2}}
\frac{ r_{\parallel}^2}{ r_{\perp}} \right)
\EEQ
where $\Psi(a,x)$ is defined by the series
\BEQ
\Psi(a,x)=\Xi(2,a;x) = \sum_{k=0}^{\infty}{(-1)^k\over k!}
{\Gamma\left({k\over 2}+a\right)\over
\Gamma\left({k\over 2}+{3\over 4}\right)}x^k \label{ser1}
\EEQ
and $a$ is defined in eq.~(\ref{eq:adef}).
Because $d$ and $m$ are integers, we have $a=n/4$
where $n$ is a positive integer. It is sufficient to distinguish
between the four cases $n=1,2,3,4$, which we shall examine below.
Indeed, the other cases
can be easily found from the recursion relation eq.~(\ref{XiRekurs}). In
figure~1, we display the normalized scaling functions $\Psi(a,x)/\Psi(a,0)$
for the cases we now proceed to study.

\subsection{$\bf a={1\over 4}$}

For this case we separate our absolutely convergent
series $\Psi(a,x)$ in two series
for the odd and even terms and use eq.~(\ref{factor}) (with $L=1$)
\BEA
\Psi\left({1\over 4},x\right)
& = & \sum_{k=0}^{\infty}{x^{2k}\over (2k)!}
{\Gamma\left(k+{1\over 4}\right)\over
\Gamma\left(k+{3\over 4}\right)} -x
\sum_{k=0}^{\infty}{x^{2k}\over (2k+1)!}{\Gamma\left(k+{3\over 4}\right)\over
\Gamma\left(k+{5\over 4}\right)}\nonumber \\
& = & \sqrt{\pi}\left[ \sum_{k=0}^{\infty}
{1\over k!}{\Gamma\left(k+{1\over 4}\right)\over
\Gamma\left(k+{1\over 2}\right)\Gamma\left(k+{3\over 4}\right)}
\left({x\over 2}\right)^{2k}\right.\nonumber \\
& & \mbox{}\qquad \left. -{x\over 2}
\sum_{k=0}^{\infty}{1\over k!}{\Gamma\left(k+{3\over 4}\right)\over
\Gamma\left(k+{5\over 4}\right)\Gamma\left(k+{3\over 2}\right)}
\left({x\over 2}\right)^{2k}\right]
\EEA
Then we use the identity eq.~(10.37.7) from \cite{hansen}
\BEQ
\sum_{k=0}^{\infty}{\Gamma\left(k+a\right)\over
k! \Gamma\left(k+2a\right)\Gamma\left(k+a+{1\over 2}\right)}
x^{2k}=\sqrt{\pi} x^{1-2a}\hbox{I}^2_{a-{1\over 2}}(x)
\EEQ
and find
\BEQ
\Psi\left( \frac{1}{4},x\right)=\sum_{k=0}^{\infty}{(-1)^k\over k!}
{\Gamma\left({k\over 2}+{1\over 4}\right)\over
\Gamma\left({k\over 2}+{3\over 4}\right)}x^k=x^{{1\over 2}}
\left[\hbox{I}_{-{1\over 4}}\left({x\over 2}\right)+
\hbox{I}_{{1\over 4}}\left({x\over 2}\right)\right]
\hbox{K}_{{1\over 4}}\left({x\over 2}\right)
\EEQ
where $\hbox{I}_{\nu}$ and $ \hbox{K}_{\nu}$ are modified Bessel functions.
This gives the exact correlation function.
Using the known asymptotic form of the Bessel functions \cite{abr},
the asymptotic behaviour is, as $x\rar\infty$
\BEQ
\Psi\left(\frac{1}{4},x\right) =
\sum_{k=0}^{\infty}{(-1)^k\over k!}{\Gamma\left({k\over 2}+{1\over 4}\right)
\over\Gamma\left({k\over 2}+{3\over 4}\right)}x^k\simeq
{2\over \sqrt{x}}\left\{1+{\cal O}\left({1\over x^2}\right)\right\}
\EEQ

\subsection{$\bf a={1\over 2}$}

We again split the expression for $\Psi(a,x)$ as before and get
\BEA
\Psi\left(\frac{1}{2}, x\right) &=& \sqrt{\pi} \left[ \sum_{k=0}^{\infty}
\frac{1}{k! \Gamma\left( k + \frac{3}{4} \right)} \left( \frac{x}{2}
\right)^{2k} \right. \nonumber \\
& & \mbox{}\qquad\left.
-\frac{x}{2}\sum_{k=0}^{\infty}\frac{1}{\Gamma\left(k+{3 \over 2} \right)
\Gamma\left( k + {5 \over 4}\right) } \left( \frac{x}{2}\right)^{2k} \right]
\EEA
Then we recall the identities eqs.~(10.7.11) and (10.7.18) from \cite{hansen}
\BEA
\sum_{k=0}^{\infty}{1\over k!\Gamma(k+a)}x^{2k}& = &
x^{1-a}\hbox{I}_{a-1}(2x) \nonumber \\
\sum_{k=1}^{\infty}{1\over \Gamma\left(k+{1\over 2}\right)\Gamma(k+a)}x^{2k}&=&
x^{{3\over 2}-a}\hbox{\bf L}_{a-{1\over 2}}(2x)
\EEA
where $\hbox{\bf L}_{\nu}(x)$ is a modified Struve function.
The asymptotic behaviour of this is given in eq.~(12.2.6) of \cite{abr} for
$x\rar\infty$ and we finally obtain
\BEA
\Psi\left({1\over 2},x\right)
& = & \sqrt{\pi}\left({x\over 2}\right)^{1\over 4}
\left[\hbox{I}_{-{1\over 4}}(x)-\hbox{\bf L}_{-{1\over 4}}(x)
\right] \nonumber \\
& \simeq & {2 \over \Gamma\left({1\over 4}\right)}
{1\over x}\left\{1+{\cal O}\left({1\over x^2}\right)\right\}
\EEA

\subsection{$\bf a={3\over 4}$}

For this particular case the series reduces to an exponential
\BEQ
\Psi\left({3\over 4},x\right)=\hbox{e}^{-x}
\EEQ

\subsection{$\bf a=1$}

In this case the model is at its upper critical dimension.
The calculation proceeds along the same lines as above and with the same
relations as for the case $a={1\over 2}$ we find

\BEA
\Psi\left(1,x\right)
& = & \sqrt{\pi}\left({1\over \sqrt{\pi}
\Gamma\left({3\over 4}\right)}+\left({x\over 2}\right)^{3\over 4}
\left[\hbox{\bf L}_{{1\over 4}}(x)
-\hbox{I}_{{1\over 4}}(x)\right]\right)\nonumber \\
& \simeq & -{1 \over 2\Gamma\left({3\over 4}\right)}
{1\over x^2}\left\{1+{\cal O}\left({1\over x^2}\right)\right\}
\EEA
where the analytic continuation $\Gamma (-1/4) = -4 \Gamma (3/4)$ was used.
We note that in this case the correlations show a
predominantly antiferromagnetic behaviour. In particular, since $\Psi(1,0)> 0$,
this implies that there exists some $x_0$ such that $\Psi(1,x_0)=0$, that is,
the universal part of the correlation function vanishes. A numerical
calculation
yields $x_0 \simeq 2.80187\ldots$ and from figure~1, it can be seen that this
is the only zero of $\Psi(1,x)$ for $x$ positive.

\section{Lifshitz points of arbitrary order}

Going beyond the simplest case $L=2$, we could attempt to repeat the
approach of the last section. In fact, we may write for any $L$
the scaling function in terms
of the generalized hypergeometric function $_{p}F_{q}$. For $L=3$, this leads
to
\BEA
\lefteqn{ \Xi\left(3,a;x\right) =
\frac{\Gamma(a)}{\sqrt{\pi}\Gamma(5/6)}
{_{1}F_{4}}\left( a; \frac{1}{3},\frac{1}{2},\frac{2}{3},\frac{5}{6};
-\frac{x^3}{27} \right) } \\
&& \!\!\!\!\! -3x \frac{\Gamma\left(a + \frac{1}{3}\right) }{2\pi}
{_{1}F_{4}}\left( a+\frac{1}{3} ;
\frac{2}{3},\frac{5}{6},\frac{7}{6},\frac{4}{3};
-\frac{x^3}{27}\right)
+ 6x^2\frac{\Gamma\left(a+\frac{2}{3}\right)}{\sqrt{\pi}\Gamma(1/6)}
{_{1}F_{4}}\left( a+\frac{2}{3} ;
\frac{7}{6},\frac{4}{3},\frac{3}{2},\frac{5}{3};
-\frac{x^3}{27}\right)
\nonumber
\EEA
A much simpler form can be obtained for the two special cases
\BEQ
\Xi \left(3,\frac{1}{2};x \right) =
\sum_{k=0}^{\infty} \frac{ (-x)^{k}}{ k! \Gamma\left( \frac{k}{3} +
\frac{5}{6} \right) } = \left( 3888 \pi^3 \right)^{1/6}
{\rm Ai}\left( - \sqrt[6]{12} x^{1/2}\right)
{\rm Ai}\left( \sqrt[6]{12} x^{1/2}\right)
\EEQ
and
\BEA
\lefteqn{ \Xi\left( 3,\frac{5}{6};x \right) =
\sum_{k=0}^{\infty} \frac{ (-x)^{k}}{k! \Gamma\left( \frac{k}{3} +
\frac{1}{2}\right) } } \\
&=& -\pi^{3/2} \left[ {\rm Ai}\left(-\sqrt[6]{12} x^{1/2}\right)
{\rm Ai}'\left(\sqrt[6]{12} x^{1/2}\right) +
{\rm Ai}'\left(-\sqrt[6]{12} x^{1/2}\right)
{\rm Ai}\left(\sqrt[6]{12} x^{1/2}\right) \right] \nonumber
\EEA
where ${\rm Ai}(x)$ is the Airy function and the prime denotes the derivative.
For the proof of these see appendix B. We shall see below that these two cases
are rather distinctive in their asymptotic behaviour for large values of $x$.
We display the scaling functions for $L=3$ in
figure~2. As we have already noted for
the case $L=2$ above, we may have
either long-range ferromagnetic or antiferromagnetic
behaviour. In distinction to the examples seen so far, in these two cases
it is known from the properties of ${\rm Ai}(x)$ that
there are infinitely many
zeroes of the scaling functions in the cases $a=1/2$ and $a=5/6$.

While these examples use some peculiarities for a given value of $L$ or $a$,
we now examine the asymptotic behaviour as $x\rar\infty$.
This follows from a general theorem due to Wright \cite{wright} on the
asymptotic behaviour of an extension of the generalized hypergeometric
function. We summarize those of his results relevant for
us in appendix A.
As it has been shown for the Lifshitz point of first order in the previous
section, the form of the asymptotic behaviour of $\Xi(L,a;x)$ depends quite
sensitively on the value of $a$.

\subsection{Algebraic asymptotic behaviour}

We first consider the case of generic values of $0 < a \leq 1$. Then from
theorem~1 of appendix A \cite{wright}, we see that the asymptotic behaviour of
the function $\Xi(L,a;x)$ is given by the poles of
the coefficients of its series
expansion eq.~(\ref{eq:XiReihe}). For generic values of
$a$, the $\Gamma$-function
in the numerator will not cancel with one of those in the denominator.
Working out the residues at those points where $\Gamma(k/L +a)$
has a pole, we find the asymptotic behaviour of
$\Xi(L,a;x)$ for $x\rar\infty$ to be
algebraic and given by
\BEQ \label{eq:AsyPot}
\Xi\left(L,\frac{d-d_{-}}{2};x\right)
\simeq \sum_{l\geq0} P_{l} \, x^{-L\left(l+{d-d_-\over 2}\right)}
\EEQ
with
\BEQ
P_l={(-1)^l \over \Gamma(l+1)}
{L\Gamma\left(L\left(l+{d-d_-\over 2}\right)\right)
\over \prod_{n=1}^{L-1}
\Gamma\left(-l-{d-d_-\over 2}+{n\over L}+{1\over 2L}\right)}
\EEQ
The reader may compare this general form with the specific results for $L=2$
found in section~3.

Since the leading term is given by $l=0$, let us consider $P_0$. We first note
that there will be a cancellation of factors in eq.~(\ref{eq:XiReihe}) if
\BEQ \label{eq:anDef}
a = a_{n} := \frac{2n+1}{2L} \;\; ; \;\; n = 1, \ldots , L-1
\EEQ
and we take the convention that $a_0 :=0$.
In fact, if $a=a_{n}$, we have to reconsider the asymptotic
behaviour and we shall
do so below. However, we note that $P_{0}$ changes sign if we pass from
$a< a_{n}$ to $a > a_{n}$ for some $n=1,\ldots L-1$. We therefore see that
\BEA
P_{0} &>0& ; \;\; {\rm if}~ a_n < a < a_{n+1}
{}~\mbox{\rm with $n$ even} \nonumber \\
P_{0} &<0& ; \;\; {\rm if}~ a_n < a < a_{n+1}
{}~\mbox{\rm with $n$ odd} \label{eq:PlusMinus}
\EEA
{}From table~1, it is now easy to see for which values of $d$ and $m$ the
leading
long-range behaviour of the spin-spin correlation function
will be ferromagnetic or antiferromagnetic, respectively.
In fact, it is quite surprising to see that already at the Lifshitz point the
anisotropies of the model can become so strong that the ferromagnetic behaviour
may be changed into an effective antiferromagnetic behaviour.
In those cases, where the system is
antiferromagnetic for large values of $x$, it
follows that the universal scaling function
vanishes for some finite $x_0$, since $\Xi(L,a;0) > 0$.
If $x=x_0$, one phenomenologically
observes effective exponents different from those
quoted in the introduction. In fact,
the mechanism of modifying scaling relations of
apparent exponents by the vanishing
of the scaling function (in distinction to the presence of
dangerous irrelevant variables)
well below the upper critical dimension appears to
occur quite generally, see \cite{Henk90}. However,
as can be seen from figure~2, even
in cases where $P_0 > 0$, this does not necessarily imply that the correlations
are ferromagnetic for all values of $x$.
An example is provided by $\Xi(3,1;x)$ which has
two zeroes at $x_{0}^{(1)} \simeq 1.231\ldots$
and $x_{0}^{(2)} \simeq 5.116\ldots$,
see figure~2. The correlations are antiferromagnetic
for $x_{0}^{(1)} < x < x_{0}^{(2)}$.
Ferromagnetic correlations appear to be kept for all $x$,
however, if $a < a_{1}$.

\subsection{Exponential-like asymptotic behaviour}

The case where $a=a_{n}$ with $a_n$ given by eq.~(\ref{eq:anDef}) marks just
the
borderline between regions of long-range ferromagnetic and antiferromagnetic
behaviour. If $1 \leq n \leq L-1$, two
$\Gamma$-functions in eq.~(\ref{eq:XiReihe}) with the same argument
cancel and we get
\BEQ
\Xi(L,a_{n};x)=\sum_{k=0}^{\infty}{(-1)^k
x^k \over k! \prod_{\ell=1}^{n-1}\Gamma\left({k\over L}
+{1\over 2L}+{\ell\over L}
\right)
\prod_{\ell=n+1}^{L-1}\Gamma\left({k\over L}+{1\over 2L}+{\ell\over L}\right)}
\EEQ
The asymptotic behaviour of this series as $x\rar\infty$
is exponential-like as found from theorem~2 of
appendix A \cite{wright}
\BEA
\lefteqn{\Xi(L,a_n;x)  \simeq  \alpha
(2\pi)^{1-{L\over 2}}\left({2L\over L-1}\right)^{{1\over 2}}
\left({x\over L}\right)^{{L\over 2(L-1)}\left({1\over 2L}-1+{a_n}
\right)} }
\nonumber \\
 && \quad\times \exp\left[2{L-1\over L}L^{{L\over 2(L-1)}-1}x^{{L\over
2(L-1)}}
\cos\left({\pi\over 2}{L\over L-1}\right)\right] \nonumber \\
  & & \quad\times \cos\left[
2{L-1\over L}L^{{L\over 2(L-1)}-1}x^{{L\over 2(L-1)}}
\sin\left({\pi\over 2}{L\over L-1}\right) +
{\pi\over 2}{L\over L-1}\left({1\over 2L}-1+
{a_n}\right)
\right] \nonumber \\
& & \quad \times
\left\{1+{\cal O}\left(x^{-{L\over 2(L-1)}}\right)\right\}
\label{eq:AsyExp}
\EEA
where $\alpha=1/2$ if $L=2$ and $\alpha=1$ if $L>2$.
Note that for $L\geq 2$ the argument of the exponential
is always negative. We also
see the presence of an oscillating term, which is absent only for $L=2$. This
indicates the presence of an infinite set of values of $x$ for which the
scaling function will vanish.

\section{Discussion}

We have found exactly the spin-spin correlation function for the anisotropic
Lifshitz points of arbitrary order $L$ realized in the spherical model with
competing interactions extending beyond the nearest neighbors. The calculation
was performed using the scaling limit
where $r_{\|},r_{\perp} \rar\infty$ simultaneously
but such that the ratio $r_{\|}/r_{\perp}^{\theta}$ is kept fixed, where
$\theta=1/L$. The result can be generally written in the form
\BEQ \label{eq:Endergebnis}
\Gll = B_1 r_{\perp}^{-(d-d_{-})}
\Phi\left( B_2 \frac{r_{\|}}{r_{\perp}^{\theta}}\right)
\EEQ
The explicit expressions for the scaling function $\Phi$ and the non-universal
metric factors $B_{1}$ and $B_2$ are given in eq.~(\ref{eq:Skalenform}). We
have
described in sections~3 and 4 the explicit representation of $\Phi$ in terms
of well-known transcendental functions. Our results are as follows.
\begin{enumerate}
\item The general form eq.~(\ref{eq:Endergebnis}) is in agreement with the
expected anisotropic scale invariance. The scaling function $\Phi$ only
depends on the number of dimensions $d$, the number $m$ of dimensions with
competing interactions present and the order $L$ of the Lifshitz point. It is
independent, for example, of the values $\kappa_i$ and we confirm the
expected universality. Properties of the model dependent on further details
of the lattice only enter into the metric factors $B_{1,2}$. We also note that
the dependence on $m$ of $\Phi$ only enters via
the lower critical dimension $d_{-}$.
\item In general, the leading asymptotic behaviour
of $\Phi$ for large values of
its argument is given by a remarkably simple structure
\BEQ \label{eq:jenesaispas}
\Phi\left(\frac{r_{\|}}{r_{\perp}^{\theta}}\right) \simeq {\cal A}
\left( \frac{r_{\|}}{r_{\perp}^{\theta}}\right)^{-(d-d_{-})/\theta}
\EEQ
where ${\cal A}$ is a known constant, see eq.~(\ref{eq:AsyPot}).
This is consistent with the known
critical exponents. If we had known beforehand that the leading
behaviour of $\Phi(x)$ for $x$ large would be a power law,
we could have predicted eq.~(\ref{eq:jenesaispas}) from
matching the correlation function scaling forms eq.~(\ref{eq:AxScal}).
\item The scaling amplitude ${\cal A}$ may be either positive or negative,
corresponding to long-range ferromagnetic or antiferromagnetic behaviour,
respectively. It is surprising to see that already at a Lifshitz point, the
effect of the competing interactions may become so strong as to be capable
to create effective antiferromagnetic correlations. Which of the two
possible situations is realized only depends on the quantity $d-d_{-}$ as
given in eq.~(\ref{eq:PlusMinus}), since ${\cal A}$ is proportional to $P_0$.
\item In those cases where the long-range behaviour is antiferromagnetic,
there is always a particular choice $x=x_0$ of the
scaling variable such that the
universal scaling function $\Phi(x_0) = 0$. The long-range correlation is
ferromagnetic for $x<x_0$ and antiferromagnetic for $x>x_0$. Antiferromagnetic
correlations will be present for at least some
values of $x$ if $d-d_{-} > 3/L$.
\item The borderline between the long-range ferromagnetic and antiferromagnetic
behaviour occurs when $d-d_{-}= \theta (2n+1)$ with $n$ being
a positive integer and
is characterized by an exponential-like behaviour
\BEA
\Phi\left({ r_{\parallel}\over r_{\perp}^{\theta}}\right) & \simeq & \alpha
  \left({ r_{\parallel}\over r_{\perp}^{\theta}}\right)
  ^{{1\over 1-\theta}\left({\theta\over 2}-1+{d-d_-\over 2}\right)}
   \exp\left(\beta
   \left({ r_{\parallel}\over r_{\perp}^{\theta}}\right)^{{1\over 1-\theta}}
\right)
   \nonumber \\
   &   & \mbox{}\times
   \cos\left(\gamma+\delta
   \left({ r_{\parallel}\over r_{\perp}^{\theta}}\right)^{{1\over 1-\theta}}
\right)
\EEA
where $\alpha, \beta, \gamma$ and $\delta$ are known constants,
see eq.~(\ref{eq:AsyExp}). We have seen that in this case there may
occur infinitely many changes between
long-range ferromagnetic and antiferromagnetic
behaviour as the scaling variable is varied. For a Lifshitz point of
first order, no zeroes occur.
\item Our results may be considered as an analogy to the calculation of
time-delayed correlation functions in dynamical problems. The analogy with
these
works for the cases $m=1$, where the role of time is
played by $r_{\|}$ and where
the analogue of a dynamical exponent $z=1/L$, and for
$m=d-1$, where $z=L$ and the role
of time is played by $r_{\perp}$.
\item Considering the case of a dynamical exponent $z=2$, we see that indeed
for
$L=2$, $d=6$ and $m=5$, the prediction
eq.~(\ref{ff1}) following from the hypothesis of
Schr\"odinger invariance at an strongly anisotropic critical point is
reproduced. For all other $d< d_{+}$, the upper critical dimension,
the scaling function has a different form.
Since local scale invariance is central to this hypothesis,
a consideration of the
cases where $d > d_{+}$ does not appear to be of much interest in this context.
\item Considering the analogy with a dynamical
exponent $z\neq 2$, a simple pattern
emerges for the cases where the scaling function has a leading exponential-like
behaviour of the form (where $\mu$ is a non-universal constant)
\BEQ
\Phi\left(\frac{r^{z}}{t}\right) \sim
\exp\left( -\mu \left(\frac{r^{z}}{t}\right)^{1/(z-1)}
\right) \label{eq:HYP}
\EEQ
for large values of its argument, and where we
suppressed the oscillating and power-like
prefactors. If $z=L$, this case is realized for the dimensions $d=L+2(n+1)$,
$n=1,\ldots,L-1$, while for $L$ even and $z=1/L$,
this case appears only for $d=4$.
We note that the form of this result
contains the number of dimensions
only implicitly through the value of the dynamical exponent $z$.
The form eq.~(\ref{eq:HYP})
is quite distinct from the conformal invariance
prediction \cite{Cardy} of the $z$-independence of
the time-dependent correlation function.
\end{enumerate}

Summarising, we have seen that already such a simple strongly
anisotropic model like the spherical model
with competing beyond nearest neighbor
interactions such as to display Lifshitz points
has quite a complicated behaviour of its spin-spin correlation function.
The results are in agreement with scale invariance and allow for the first
time to ask questions about the form of the scaling function itself.
While in a few cases, the results can be understood in terms of local
scale invariance, it remains a challenge to develop a better
conceptual understanding of these fascinating phenomena.

\zeile{2}
\noindent{\bf Acknowledgement:} It is a pleasure to thank M. Droz for
useful comments.

\appendix
\renewcommand{\theequation}{A.\arabic{equation}}
\appsection{A}

We recall a few results on the asymptotics of the following
extension of the generalized hypergeometric function
\BEQ
{_{p}{\cal F}_{q}}(x) = \sum_{k=0}^{\infty} \frac{f(k)}{k!} (-x)^{k}
\EEQ
where
\BEQ
f(t) = \left( {\prod_{r=1}^{p} \Gamma(\alpha_r t + \beta_r)}\right)
\left( {\prod_{r=1}^{q}
\Gamma( \rho_{r} t + \mu_r )}\right)^{-1}
\EEQ
If $\alpha_r =1$ and $\rho_r =1$ for all values $r$ occurring, we recover the
generalized hypergeometric function as defined e.g. in \cite{Erde55}.
The numbers $\alpha_r$ and $\rho_r$ are all real and positive and
\BEQ
\kappa=1+\sum_{r=1}^q \rho_r-\sum_{r=1}^p \alpha_r>0
\EEQ
For our purposes, where
\BEQ
\kappa=2-{2\over L}
\EEQ
we need the asymptotic behaviour along the positive real axis
for values of $1\leq\kappa<2$. For the convenience of the reader
we restate the relevant theorems obtained by Wright \cite{wright}. The
full asymptotic expansion for any complex argument and for any
$\kappa>0$ can be found in \cite{wright}.

\noindent {\bf Theorem 1:} {\it If $0<\kappa<2$, then the asymptotic expansion
of $_p{\cal F}_q(x)$ for $x\rar\infty$ is
\BEQ
_p{\cal F}_q=J(x)
\EEQ
where $J(x)$ is defined below. }

\noindent {\bf Theorem 2:} {\it If $f(t)$ has only a finite number of poles
or none, then $\kappa\geq 1$ and the asymptotic expansion of $_p{\cal F}_q(x)$
for $x\rar\infty$ is given by
\BEA
_p{\cal F}_q(x)&=&I(Z)+I(\bar{Z})+H(x) \quad (1<\kappa<2)\, ,\nonumber\\
_p{\cal F}_q(x)&=&I(Z)+H(x) \quad (\kappa=1)
\EEA
where $I(X),H(x)$ and $Z$ are defined below.
If $f(t)$ has no poles, then $H(x)=0$. }

The following notation is used. Let
\BEQ
h=\left(\prod_{r=1}^p \alpha_r^{\alpha_r}\right)
\left(\prod_{r=1}^q \rho_r^{-\rho_r}\right),\quad
\vartheta=\sum_{r=1}^p \beta_r-\sum_{r=1}^q \mu_r+{1\over 2}(q-p)
\EEQ
The variable $Z$ is defined as
\BEQ
Z=\kappa(hx)^{1/ \kappa}\hbox{e}^{i{\pi/ \kappa}}
\EEQ
and we write $I(X)$ for the exponential asymptotic expansion
\BEQ
I(X)=A_0X^{\vartheta}\hbox{e}^X\left[1+{\cal O}(X^{-1})\right]
\EEQ
where
\BEQ
A_0=(2\pi)^{{1\over 2}(p-q)}\kappa^{-{1\over 2}-\vartheta}
\prod_{r=1}^p\alpha_r^{\beta_r-{1\over 2}}
\prod_{r=1}^q\rho_r^{{1\over 2}-\mu_r}\, ,
\EEQ
and $J(y)$ for the algebraic expansion
\BEQ
J(y)=\sum_{r=1}^p \sum_{l\geq 0}P_{r,l}y^{-{(l+\beta_r)/ \alpha_r}}
\EEQ
where $P_{r,l}$ are defined from the poles of $f(t)$ in the following way.
The poles of $f(t)$ are among those of
$\prod_{r=1}^p\Gamma(\alpha_rt+\beta_r)$
at the points
\BEQ
t=-{l+\beta_r\over \alpha_r}\, .
\EEQ
If $f(t)$ has a pole of degree $s$ at this point, we write for the residue
\BEQ
sP_{r,l}y^{-{(l+\beta_r)/ \alpha_r}}=\hbox{Res} \left(\Gamma(-t)f(t)y^t \right)
\EEQ

If $f(t)$ has only a finite number $l_r$ of poles, then $P_{r,l}=0$
when $l> l_r$ and $H(y)$ is the finite sum
\BEQ
H(y)=\sum_{r=1}^p \sum_{l=0}^{l_r}P_{r,l}y^{-{l+\beta_r\over \alpha_r}}
\EEQ

\renewcommand{\theequation}{B.\arabic{equation}}
\appsection{B}

We analyse a few sums arising in the calculation of some of the
scaling functions discussed in the text. We begin with
\BEQ
\psi(x) := \Xi\left( 3,\frac{1}{2};x\right) =
\sum_{k=0}^{\infty} \frac{(-x)^{k}}{k!
\Gamma\left(\frac{k}{3}+\frac{5}{6}\right)}
\EEQ
where $x$ is real and positive.
To bring this into a more tractable form, we use the identity
\BEQ \label{Fak3}
(3k)! = (2\pi)^{-1} 3^{3k+\frac{1}{2}} k! \Gamma\left(k+\frac{1}{3}\right)
\Gamma\left(k+\frac{2}{3}\right)
\EEQ
and get $\psi = \psi_1 + \psi_2 + \psi_3$, where each term is treated
separately. The first one is
\BEA
\psi_1 &=& \sum_{k=0}^{\infty}
\frac{(-x)^{3k}}{(3k)! \Gamma(k + 5/6)} \nonumber \\
&=& \frac{2^{7/6}\pi}{\sqrt{6\pi}} \sum_{k=0}^{\infty} \frac{(-1)^k
\left( \frac{4}{27} x^3 \right)^{k}}{k! \Gamma(k+2/3)\Gamma(2k+2/3)}
\EEA
where we reused eq.~(\ref{factor}).
Next, we recall the identity eq.~(10.40.2) from \cite{hansen}
\BEQ \label{Id:Hansen2}
\sum_{k=0}^{\infty} \frac{(-1)^{k} x^{4k}}{k! \Gamma(k+a)\Gamma(2k + a)}
= x^{2-2a}\FB{I}{a-1}{2x}\FB{J}{a-1}{2x}
\EEQ
where $\FB{I}{\nu}{x}$ and $\FB{J}{\nu}{x}$ are Bessel functions, and get
\BEQ
\psi_1 = {2\over 3}\sqrt{\pi x}
\FB{I}{-\frac{1}{3}}{X}\FB{J}{-\frac{1}{3}}{X}
\EEQ
with the abbreviation
\BEQ
X := \left(\frac{64}{27}\right)^{1/4} x^{3/4}
\EEQ
The second term is treated in an analogous fashion and we have
\BEA
\psi_2 &=& \sum_{k=0}^{\infty}
\frac{(-x)^{3k+1}}{(3k+1)! \Gamma(k+7/6)} \nonumber \\
&=& - \frac{2^{11/6}\pi}{\sqrt{6\pi}} \frac{x}{3} \sum_{k=0}^{\infty}
\frac{(-1)^k \left(\frac{4}{27}x^3\right)^{k}}{k! \Gamma(k+4/3)
\Gamma(2k + 4/3)} \nonumber \\
&=& - {2\over 3}\sqrt{\pi x}
\FB{I}{\frac{1}{3}}{X}\FB{J}{\frac{1}{3}}{X}
\EEA
The third term finally is
\BEA
\psi_3 &=& \sum_{k=0}^{\infty}
\frac{(-x)^{3k+2}}{(3k+2)! \Gamma(k+3/2)} \nonumber \\
 &=& \frac{2\pi}{\sqrt{3}} \left(\frac{x}{3}\right)^{2}
\sum_{k=0}^{\infty} \frac{(-x/3)^{3k}}{k!
\Gamma(k+3/2)\Gamma(k+4/3)\Gamma(k+5/3)} \nonumber \\
 &=& {2\over 3}\sqrt{\pi x}  \sum_{k=0}^{\infty}
\frac{(-1)^k \left[ \left(\frac{4}{27}\right)^{1/4}
x^{3/4}\right]^{4k+2}}{(2k+1)! \Gamma(k+4/3) \Gamma(k+5/3)}
\EEA
We now use the identity eq.~(10.40.8) from \cite{hansen}
\BEA
\lefteqn{\sum_{k=0}^{\infty}
\frac{(-1)^{k} x^{4k+2}}{(2k+1)! \Gamma(k+3/2 -a)
\Gamma(k+3/2 + a)} }\nonumber \\
&=&  \frac{1}{2\sin \pi a} \left[ \FB{J}{2a}{2x}\FB{I}{-2a}{2x}
-\FB{J}{-2a}{2x}\FB{I}{2a}{2x} \right] \label{Id:Hansen1}
\EEA
and find
\BEQ
\psi_3 = {2\over 3}\sqrt{\pi x}
\left[ \FB{J}{\frac{1}{3}}{X}\FB{I}{-\frac{1}{3}}{X} -
\FB{J}{-\frac{1}{3}}{X}\FB{I}{\frac{1}{3}}{X} \right]
\EEQ
We collect the results and get
\BEA
\psi(x) &=& {2\over 3}\sqrt{\pi x}  \left[
\FB{J}{-\frac{1}{3}}{X}\FB{I}{-\frac{1}{3}}{X}
- \FB{J}{\frac{1}{3}}{X}\FB{I}{\frac{1}{3}}{X} +
\FB{J}{\frac{1}{3}}{X}\FB{I}{-\frac{1}{3}}{X} -
\FB{J}{-\frac{1}{3}}{X}\FB{I}{\frac{1}{3}}{X} \right] \nonumber \\
&=& {2\over 3}\sqrt{\pi x}  \left[
\FB{J}{\frac{1}{3}}{X}+\FB{J}{-\frac{1}{3}}{X}\right]
\cdot\left[\FB{I}{-\frac{1}{3}}{X}-\FB{I}{\frac{1}{3}}{X}\right] \nonumber \\
&=& \left( 3888 \pi^3\right)^{1/6} {\rm Ai}\left( -\sqrt[6]{12} x^{1/2}\right)
{\rm Ai}\left( \sqrt[6]{12} x^{1/2}\right)
\EEA
where ${\rm Ai}(x)$ is the Airy function.
This is the result quoted in the text.

We next consider the function
\BEQ
\vph(x) := \Xi \left( 3,\frac{5}{6};x \right) =
\sum_{k=0}^{\infty} \frac{(-x)^k }{k!
\Gamma\left( \frac{k}{3}+\frac{1}{2}\right)}
\EEQ
We use again eq.~(\ref{Fak3}) to get the decomposition
$\vph = \vph_1 + \vph_2 + \vph_3$ and turn to study these terms separately.
The first one is
\BEA
\vph_1 &=& \sum_{k=0}^{\infty}
\frac{(-x)^{3k}}{(3k)! \Gamma(k+1/2)} \nonumber \\
&=& 2\sqrt{\frac{\pi}{3}} \sum_{k=0}^{\infty}
\frac{(-1)^k \left(\frac{4}{27} x^3
\right)^{k}}{(2k)! \Gamma(k+1/3) \Gamma(k+2/3)}
\EEA
This is rewritten via the identity
\BEA
\lefteqn{ \sum_{k=0}^{\infty} \frac{(-1)^k x^{4k}}{(2k)! \Gamma(k+1/2-a)
\Gamma(k+1/2 +a)} } \label{iden1} \\
&=& -\frac{x}{4\sin a\pi}
\left\{ \FB{J}{2a}{2x}\FB{I}{-2a-1}{2x}
-\FB{J}{-2a}{2x}\FB{I}{2a-1}{2x} \right.
\nonumber \\ & &\mbox{}\quad \left.
-\FB{J}{-2a+1}{2x}\FB{I}{2a}{2x}+
\FB{J}{2a+1}{2x}\FB{I}{-2a}{2x}\right\} \nonumber
\EEA
We postpone the proof of this and proceed with the calculation. We get
\BEQ
\vph_1 = -\sqrt{\frac{\pi}{12}} X
\left\{ \FB{J}{\frac{1}{3}}{X}\FB{I}{-\frac{4}{3}}{X}
-\FB{J}{-\frac{1}{3}}{X}\FB{I}{-\frac{2}{3}}{X}-
\FB{J}{\frac{2}{3}}{X}\FB{I}{-\frac{1}{3}}{X}
+\FB{J}{\frac{4}{3}}{X}\FB{I}{-\frac{1}{3}}{X}\right\}
\EEQ
For the second term we have
\BEA
\vph_2 &=& \sum_{k=0}^{\infty} \frac{(-x)^{3k+1}}{(3k+1)!
\Gamma(k+5/6)} \nonumber \\
&=& - 2^{5/3}\sqrt{\frac{\pi}{3}}\frac{x}{3}
\sum_{k=0}^{\infty} \frac{(-1)^{k} \left(\frac{4}{27} x^3\right)^{k}}{k!
\Gamma(k+2/3)\Gamma(2k+5/3)}
\EEA
This is evaluated by using the identity
\BEQ \label{iden2}
\sum_{k=0}^{\infty} \frac{(-1)^k x^{4k}}{k! \Gamma(k+a)\Gamma(2k+a+1)}
= \frac{x^{1-2a}}{2} \left[ \FB{I}{a-1}{2x}\FB{J}{a}{2x}+
\FB{I}{a}{2x}\FB{J}{a-1}{2x} \right]
\EEQ
which we prove below. We find
\BEQ
\vph_2 = -\sqrt{\frac{\pi}{12}} X
\left[ \FB{I}{-\frac{1}{3}}{X}\FB{J}{\frac{2}{3}}{X}+
\FB{I}{\frac{2}{3}}{X}\FB{J}{-\frac{1}{3}}{X}\right]
\EEQ
and for the last term we get, again using eq.~(\ref{iden2})
\BEA
\vph_3 &=& \sum_{k=0}^{\infty}
\frac{(-x)^{3k+2}}{(3k+2)!\Gamma(k+7/6)} \nonumber \\
&=& 2^{7/3}\sqrt{\frac{\pi}{3}}\left(\frac{x}{3}\right)^{2}
\sum_{k=0}^{\infty} \frac{(-1)^{k} \left(\frac{4}{27} x^3 \right)^{k} }{k!
\Gamma(k+4/3)\Gamma(2k+7/3)} \nonumber \\
&=& \sqrt{\frac{\pi}{12}} X
\left[ \FB{I}{\frac{1}{3}}{X} \FB{J}{\frac{4}{3}}{X}+
\FB{I}{\frac{4}{3}}{X} \FB{J}{\frac{1}{3}}{X} \right]
\EEA
Combining these three terms, we find
\BEA
\vph(x) &=& \sqrt{\frac{\pi}{12}} X \left\{ \FB{J}{\frac{1}{3}}{X}\left[
\FB{I}{\frac{4}{3}}{X}-\FB{I}{-\frac{4}{3}}{X}\right] +
\FB{J}{-\frac{1}{3}}{X}\left[\FB{I}{-\frac{2}{3}}{X}-
\FB{I}{\frac{2}{3}}{X}\right] \right. \nonumber \\
& & \mbox{}\qquad\left. +\FB{J}{\frac{2}{3}}{X}\left[\FB{I}{\frac{1}{3}}{X}
-\FB{I}{-\frac{1}{3}}{X}\right]
+\FB{J}{\frac{4}{3}}{X}\left[\FB{I}{\frac{1}{3}}{X}
-\FB{I}{\frac{1}{3}}{X}\right]\right\}  \\
&=& \sqrt{\frac{\pi}{12}} X \left\{ \left[
\FB{J}{\frac{1}{3}}{X} + \FB{J}{-\frac{1}{3}}{X} \right]
\FB{K}{\frac{2}{3}}{X} + \left[
\FB{J}{-\frac{2}{3}}{X} - \FB{J}{\frac{2}{3}}{X}\right]
\FB{K}{\frac{1}{3}}{X} \right\} \nonumber
\EEA
recalling the familiar relationship between the modified Bessel functions
$\FB{I}{\nu}{X}$ and $\FB{K}{\nu}{X}$ and using
the recursions eq.~(\ref{BesRek}) below and
\BEQ
\FB{K}{\nu+1}{x} = \FB{K}{\nu-1}{x} + \frac{2\nu}{x}\FB{K}{\nu}{x}
\EEQ
Using the relationship with the Airy function ${\rm Ai}(x)$ and its derivative
the final result is
\BEQ
\vph(x) = -\pi^{3/2} \left[ {\rm Ai}\left(-\sqrt[6]{12} x^{1/2}\right)
{\rm Ai}'\left(\sqrt[6]{12} x^{1/2}\right) +
{\rm Ai}'\left(-\sqrt[6]{12} x^{1/2}\right)
{\rm Ai}\left(\sqrt[6]{12} x^{1/2}\right) \right]
\EEQ
which is the form stated in the text. In the same way, the representation of
$\Xi(3,a;x)$ in terms of the generalized hypergeometric function $_{1}F_{4}$
can
be obtained.

We now prove the identities needed in the calculation.
For the proof of eq.~(\ref{iden2}), let
\BEA
T &:=& \sum_{k=0}^{\infty} \frac{(-1)^{k} x^{4k}}{k!
\Gamma(k+a)\Gamma(2k+a+1)} \nonumber \\
 &=& \sum_{k=0}^{\infty} \frac{ (k+a) (-1)^{k}
x^{4k}}{k! \Gamma(k+a+1)\Gamma(2k+a+1)} \nonumber \\
 &=& \left( a + \frac{x}{4}\frac{\partial}{\partial x}\right)
\left( x^{-2a} \FB{I}{a}{2x} \FB{J}{a}{2x}\right)
\EEA
where eq.~(\ref{Id:Hansen2}) was used. We then use
\BEQ
\frac{d}{dx} \left( x^{-a} \FB{J}{a}{x}\right) =
- x^{-a}\FB{J}{a+1}{x} \;\; , \;\;
\frac{d}{dx} \left( x^{-a} \FB{I}{a}{x}\right) =  x^{-a}\FB{I}{a+1}{x}
\EEQ
and find
\BEQ
T = \frac{x^{1-2a}}{2} \left[
\left( \frac{a}{x} \FB{I}{a}{2x} + \FB{I}{a+1}{2x}\right)\FB{J}{a}{2x} +
\left( \frac{a}{x} \FB{J}{a}{2x} -
\FB{J}{a+1}{2x}\right)\FB{I}{a}{2x} \right]
\EEQ
Then eq.~(\ref{iden2}) follows from the
recursion relations of the Bessel functions
\BEQ \label{BesRek}
\FB{I}{\nu-1}{x} = \frac{2\nu}{x}\FB{I}{\nu}{x} + \FB{I}{\nu+1}{x} \;\; , \;\;
\FB{J}{\nu-1}{x} = \frac{2\nu}{x}\FB{J}{\nu}{x} - \FB{J}{\nu+1}{x}
\EEQ

For the proof of eq.~(\ref{iden1}), let
\BEQ
S := \sum_{k=0}^{\infty} \frac{(-1)^{k} x^{4k}}{(2k)! \Gamma(k+1/2-a)
\Gamma(k+1/2+a)}
\EEQ
We separate off the term with $k=0$. For the remaining sum,
we make a shift in the
summation index and have the decomposition $S=S_0 + S_1$ where
\BEA
S_0 &=& \frac{\cos a \pi}{\pi} \\
S_1 &=& -\frac{1}{2} \sum_{k=0}^{\infty} \frac{1}{k+1}
\frac{(-1)^{k} x^{4k+4}}{(2k+1)! \Gamma(k+3/2-a)\Gamma(k+3/2+a)} \nonumber \\
 &=& -2 \int_{0}^{x} dy \sum_{k=0}^{\infty}
\frac{(-1)^{k} y^{4k+3}}{(2k+1)! \Gamma(k+3/2-a)\Gamma(k+3/2+a)} \nonumber \\
 &=& -\frac{1}{4\sin a\pi} \int_{0}^{2x} \!du \,
u \left[ \FB{J}{2a}{u}\FB{I}{-2a}{u}
- \FB{J}{-2a}{u}\FB{I}{2a}{u} \right] \label{eq:B29}
\EEA
where the identity eq.~(\ref{Id:Hansen1}) was used.
To evaluate this, we use the following, see eq.~(11.3.29) in \cite{abr}
\BEA
\int_{0}^{x} \!du \, u \FB{J}{\nu}{k u}\FB{J}{-\nu}{l u} &=& \frac{x}{2}
\left[ k \FB{J}{\nu+1}{k x}\FB{J}{-\nu}{l x} - l
\FB{J}{\nu}{kx}\FB{J}{-\nu+1}{lx}\right]
\nonumber \\
& & -\nu \FB{J}{\nu}{kx}\FB{J}{-\nu}{lx} + \lim_{\eps\rar 0} \nu
\FB{J}{\nu}{k\eps}\FB{J}{-\nu}{l\eps}
\EEA
We now use the relation with the modified Bessel function
\BEQ
\FB{I}{\nu}{x} = \exp\left(\frac{-\nu \pi i}{2}\right)\FB{J}{\nu}{ix}
\EEQ
and take $k=1$ and $l=i$. With the leading behaviour of the Bessel functions
for
small values of their arguments \cite{abr} and some
algebra, we obtain the identity
\BEA
\int_{0}^{x} \!du \, u \FB{J}{\nu}{u}\FB{I}{-\nu}{u} &=& \frac{x}{2}
\left[ \FB{J}{\nu+1}{x}\FB{I}{-\nu}{x} + \FB{J}{\nu}{x}\FB{I}{-\nu+1}{x}\right]
\nonumber \\
& & -\nu \FB{J}{\nu}{x}\FB{I}{-\nu}{x} + \frac{\sin \pi \nu}{\pi}
\EEA
Insertion into eq.~(\ref{eq:B29}) then yields the following
\BEA
\lefteqn{S = -\frac{x}{4\sin \pi a} \left[ -
\frac{2a}{x}\left( \FB{J}{2a}{2x}\FB{I}{-2a}{2x}
+\FB{J}{-2a}{2x}\FB{I}{2a}{2x}\right) \right.} \\
&+& \left. \FB{J}{2a+1}{2x}\FB{I}{-2a}{2x}+\FB{J}{2a}{2x}\FB{I}{-2a+1}{2x}
-\FB{J}{-2a+1}{2x}\FB{I}{2a}{2x}-
\FB{J}{-2a}{2x}\FB{I}{2a+1}{2x}\frac{ }{}\right] \nonumber
\EEA
and eq.~(\ref{iden1}) is obtained with the help of the recursion
relation eq.~(\ref{BesRek}). This completes the proof.

\newpage

\zeile{1}
\noindent{\bf Figure captions}
\zeile{3}
\noindent{\bf Figure 1:} Normalized scaling function
$\Psi(a,x)/\Psi(a,0)$ for the
values $a=1/4,1/2,3/4$ and $a=1$ as a function of $x$.
\zeile{1}
\noindent{\bf Figure 2:} Normalized scaling functions
$\Xi(3,a;x)/\Xi(3,a;0)$ for
$a=1/6,1/3,1/2,2/3$, $5/6$ and $a=1$ as a function of $x$.

\end{document}